\documentstyle[preprint,aps,epsfig]{revtex}
\tighten
\begin{document}
\preprint{\vbox{Submitted to Physical Review C}}
\newcommand{\be}{\begin{equation}}
\newcommand{\ee}{\end{equation}}
\newcommand{\bea}{\begin{eqnarray}}
\newcommand{\eea}{\end{eqnarray}}
\newcommand{\bc}{\begin{center}}
\newcommand{\ec}{\end{center}}

\title{Modeling the strangeness content of hadronic matter}
\author{G. Toledo S\'anchez and J. Piekarewicz}
\address{Department of Physics,
         Florida State University, 
         Tallahassee, FL 32306-4350}
\date{\today}
\maketitle
\begin{abstract}
   The strangeness content of hadronic matter is studied in a
   string-flip model that reproduces various aspects of the
   QCD-inspired phenomenology, such as quark clustering at low
   density and color deconfinement at high density, while
   avoiding long range van der Waals forces. Hadronic matter is
   modeled in terms of its quark constituents by taking into
   account its internal flavor (u,d,s) and color (red, blue,
   green) degrees of freedom. Variational Monte-Carlo
   simulations in three spatial dimensions are performed for
   the ground-state energy of the system. The onset of the
   transition to strange matter is found to be influenced by
   weak, yet not negligible, clustering correlations. The
   phase diagram of the system displays an interesting 
   structure containing both continuous and discontinuous
   phase transitions. Strange matter is found to be absolutely 
   stable in the model.\\
\end{abstract}

PACS 12.39.-x, 24.85.+p, 26.60.+c
\section{Introduction}
\label{introduction}

Strange matter, a deconfined state of quark matter consisting of
almost equal amounts of up, down, and strange quarks, has been
speculated to be the absolute ground state of hadronic
matter~\cite{Bo71,Wi84}. If true, nucleons and nuclei --- and thus
most of the luminous matter in the universe --- is in a long-lived
metastable state. Undoubtedly, the confirmation of such hypothesis
would have far reaching consequences on a variety of fields, ranging
from astronomy and cosmology all the way to particle and nuclear
physics.  Stimulated by such an exciting possibility, searches for
strange matter are currently being conducted at terrestrial
laboratories as well as at space-based observatories.  Indeed, a
substantial effort has been devoted on experimental searches for
strangelets (``strange-matter nuggets'') at both CERN and Brookhaven
National Laboratory (BNL) and more are proposed in the future for the
relativistic heavy-ion collider (RHIC) and the large-hadronic
collider (LHC). These terrestrial experiments are being complemented
by observational searches for strange stars. What would be the
signature for such exotic objects? Since strange stars are self-bound
objects having a mass-radius relation quite different than the
gravitationally bound neutron stars, they are allowed rotational
periods considerably shorter than those predicted for gravitationally
bound stars. {\it Consequently, if a pulsar with a period falling
below the limit of gravitationally bound stars were discovered, the
conclusion that the confined hadronic phase of nucleons and nuclei is
only metastable would be virtually inescapable~\cite{GlBook97}.}

Such a pulsar may have been recently discovered~\cite{Wi98,Ch98}. 
The pulsar SAX J1808.4-3658, with a rotation period of 2.5~ms, is the 
fastest spinning x-ray pulsar ever observed. Based on a study of its 
mass-radius relation it has been concluded that SAX J1808.4-3658 is a
likely strange-star candidate~\cite{Li99}, although this interpretation
remains controversial~\cite{Gl00,Sc00}. Still, the discovery of
such a fastly-rotating pulsar appears to have made the detection of 
strange matter within observational reach. In turn, the confirmation 
of such an exotic state of matter will help settle the claim that
at present our universe is in a long-lived metastable~state.

In the present work we focus on the impact of strangeness on the
equation of state (EOS). One motivation for this study, in addition to
those mentioned earlier, is the observation that the masses of about
20 neutron stars are remarkably close to the ``canonical'' value of
$M=1.4 M_{\odot}$~\cite{TC99}. Yet conventional models of nuclear
structure, with equation of states constrained from the bulk
properties of nuclear matter, seem to allow substantially larger
masses~\cite{GlBook97,SW86}. However, the existence of a quark-matter
phase at the core of neutron stars (NS) will soften considerably the
equation of state leading to smaller limiting masses. Thus a study of
the strangeness content of hadronic matter, using a ``QCD-inspired''
model, is desirable.  For static and spherically symmetric neutron
stars obeying the Oppenheimer-Volkoff equations the only physical
ingredient that remains to be specified is the equation of state. Yet
an equation of state that is accurate over the whole range of
densities present in a neutron star remains a formidable
challenge. For example, such an equation of state should be able to
describe the hypothetical ``hybrid stars'': neutron stars consisting
of a quark-matter core below a nuclear-matter mantle.  Unfortunately,
traditional studies of strange matter has been conducted in two vastly
different pictures~\cite{Fa84,Sc93,Gr98,Ma98}.  One picture uses a
hadronic model --- similar to ordinary nuclei --- where the
fundamental degrees of freedom are mesons and baryons. The other
picture uses a quark model consisting of massless, noninteracting
quarks confined inside a bag. Presumably a description of strange
matter in terms of mesons and baryons is well motivated in the
low-density regime where clustering correlations remain important. At
the same time, strange matter viewed as a relativistic Fermi gas of
quark might be appropriate at the extremes of densities necessary for
color deconfinement to occur. Yet this division seems ad-hoc and
arbitrary; for example, at what density should one switch from a
nuclear- to quark-based description? Perhaps the most serious
difficulty encountered in modeling the density dependence of hadronic
matter and the resulting EOS is how to model a system that has quarks
confined inside color-neutral hadrons at low density but free quarks
at high density. Much of the responsibility for such complexity rests
on the self-interactions among the gluons which generate quark
confinement in the low-density regime. On the other hand, the quark
substructure of the hadrons should become important as the density
increases.

Although the evidence in support of QCD as the correct theory of the
strong interactions is overwhelming, at present no rigorous solution
of QCD exists in the regime of high-baryon density. Thus, one must
resort to QCD-inspired phenomenological models. Such models of
hadronic matter using quarks as the underlying degrees of freedom have
been developed to reproduce some properties of
QCD~\cite{Ho85,Wa89,Ho91,Ho92,Fr94}. The main feature of these models
--- generically known as ``string-flip'' models~\cite{Le86} --- is the
existence of a many-body potential able to confine quarks within
color-singlet clusters without generating unphysical long-range van
der Waals forces~\cite{Li81}. The many-body potential is evaluated by
solving a difficult optimization problem; one must decide how to
assign colored quarks into color-singlet clusters (see
below). Presumably, this ``quark-assignment'' problem is meant to
represent the optimal configuration of gluonic strings. While the
precise form of the potential is presently unknown, the many-quark
problem is likely to require solving some complicated global
optimization problem. Although string-flip models violate important
symmetries of QCD, such as chiral symmetry and Lorentz invariance,
they excel in places where most other models fail: the transition from
nuclear to quark-matter. In the string-flip moedl this transition is
dynamical without the need to rely on
ad-hoc parameters. Hence, such models
should shed light on the possibility of stable strange matter.  That
the emergence of strange quarks at high-baryon density is
energetically favorable is easy to understand. As the density of the
system increases, the Pauli exclusion principle forces the chemical
potential to increase from the light-quark mass $m$ to $E_{\rm
F}=\sqrt{k_{\rm F}^2+m^2}$, where $k_{\rm F}$ is the Fermi momentum.
What is not easy to understand are the details of the transition. For
example, do clustering correlations remain important at the transition
density or has the system evaporated into the free quarks? Does the
EOS predicted by the model yield self-bound and absolutely stable
strange stars? These are the sort of questions that we plan to address
in this paper.

An initial study of strange matter in the string-flip model was
carried out in Ref.~\cite{Mo99}. There, a highly simplified version 
of the model was used to simulate one-dimensional matter in terms of
two-color, two-flavor (``up'' and ``strange'') constituent
quarks. While it was found that clustering correlations remain
important in the transition region, strange matter was found to be
unbound. In this paper we extend the results of Ref.~\cite{Mo99} by
simulating three-flavor (up, down, and strange), three-color (``red'',
``blue'', and ``green'' ) hadronic matter in three-dimensional
space. A variety of ground-state observables are computed as a
function of density with the goal of characterizing the transition 
to strange matter and to establish the possibility of absolute
stability. We have organized the paper as follows:
Sec.~\ref{formalism} introduces the general ideas used to model a 
system of fermions focusing on the structure of the wave function
and the many-body potential. We then consider both the low- and
high-density limits to establish closed-form baseline results. 
After describing the variational Monte Carlo procedure, results
are presented for a variety of ground-state observables. Finally,
we offer conclusions and perspectives for future work in 
Sec.~\ref{conclusions}.

\section{Formalism}
\label{formalism}

\subsection{The many-quark potential}

The QCD-inspired phenomenology prescribes how to model the many-quark
potential. At very-low density the quarks must be confined within
color-singlet clusters that should interact weakly due to the
short-range nature of the nucleon-nucleon interaction. This suggests a
strong---indeed confining---force between quarks in the same cluster
but no further interaction between quarks in different clusters. Thus,
the force saturates within each color-singlet hadron.  This
saturatation is necessary in order for the hadrons to be able to
separate without generating unphysical long-range forces.  In
contrast, in the high-density domain asymptotic freedom demands that
the interaction between all quarks be weak. This behavior is expected
once the average inter-quark separation becomes smaller than the
typical confining scale. In this regime the only important correlation
among quarks will be induced by the Pauli exclusion principle and the
system should evolve into a Fermi gas of quarks.

\begin{figure}
\leavevmode\centering\psfig{file=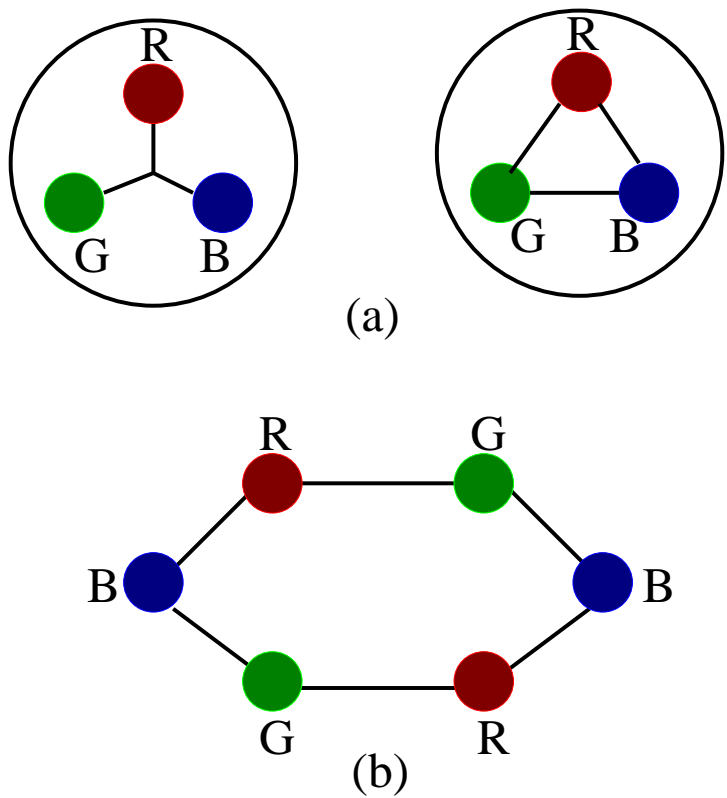,angle=00,width=4in}
\vspace{0.25in}
\caption{(a) Y-shaped (left) and triangular (right) arrangement 
	     of strings for a single three-quark cluster; for
             harmonic string the potential is identical. (b)
	     An example of a six-quark configuration allowed
             in the model.}
 \label{Figure0}
\end{figure}

A many-quark potential that meets these requirements was first
introduced by Lenz and collaborators~\cite{Le86} to model meson-meson
interactions. Soon after the potential was adapted by Horowitz, Moniz,
and Negele for the study of one-dimensional nuclear
matter~\cite{Ho85}. Several more realistic generalizations have
followed~\cite{Wa89,Ho91,Ho92,Fr94}, although all limited to
non-strange matter. It is on one of these models~\cite{Ho92} that we
base our present generalization to strange matter. The model is
constructed from quarks having flavor ({\em up, down, strange}) and
color ({\em red, blue, green}) degrees of freedom. The many-quark
potential is defined as the optimal clustering of quarks into
color-singlet objects. For reasons that will become clear later, the
implementation of this idea is carried out as follows. Consider all
red and blue quarks in the system, irrespective of flavor in
accordance to the ``flavor-blind'' nature of QCD.  We define the
``optimal pairing'' of red and blue quarks as:
\begin{equation}
  V_{RB} = \mathop{\min}_{P}
           \sum_{i=1}^{A} 
	   v\Big({\bf r}_{iR},P({\bf r}_{iB})\Big) \;,
 \label{vrb}
\end{equation}
where ${\bf r}_{iR}$ denotes the spatial coordinate of the  {\it ith} red
quark and $P({\bf r}_{iB})$ is the coordinate of the mapped {\it ith} blue 
quark 
(${\bf r}_{iB}\mapsto P({\bf r}_{iB})\!\equiv\!{\bf r}_{jB}$). 
Note that the minimization procedure is over all possible $A!$
permutations of the $A$ blue quarks and that the confining potential 
$v$ is assumed harmonic with a spring constant denoted by $k$
(see Fig.~\ref{Figure0}). That is,
\begin{equation}
  v({\bf r}_{iR},{\bf r}_{jB}) = \frac{1}{2}k 
  ({\bf r}_{iR}-{\bf r}_{jB})^{2} \;.
 \label{vstring}
\end{equation}
The ``blue-green'' and ``green-red'' components of the many-quark
potential are defined in direct analogy to Eq.~(\ref{vrb})
\begin{mathletters}
\begin{eqnarray}
  V_{BG} &=& \mathop{\min}_{P}
             \sum_{i=1}^{A} 
	     v\Big({\bf r}_{iB},P({\bf r}_{iG})\Big) \;, 
  \label{vbg} \\
  V_{GR} &=& \mathop{\min}_{P}
             \sum_{i=1}^{A} 
	     v\Big({\bf r}_{iG},P({\bf r}_{iR})\Big) \;.
  \label{vgr} 
\end{eqnarray}
\end{mathletters}
In this manner the many-quark potential to be used in our simulations
of strange matter becomes equal to:
\begin{equation}
  V = V_{RB} + V_{BG} + V_{GR} 
 \label{MQpotential}
\end{equation}

and thus, the Hamiltonian describing the system of $N$ particles each with mass $m_i$ and momentum ${\bf p}_i$ is given by:

\begin{equation}
H = \sum_{i=1}^{N}\frac{{\bf p}_{i}^{2}}{2m_i}  + V \;. 
 \label{HamiltonV}
\end{equation}

Several comments are now in order. First, the constructed potential is
able to confine quarks within color-singlet clusters. Yet the strong
confining force saturates within each color-singlet cluster allowing
the clusters to separate without generating long-range van der Waals
forces. Moreover, the potential is symmetric under the exchange of
identical quarks. Second, the potential is truly many-body as moving
one single quark might cause many of the ``strings'' to flip; note
that even those strings that are not connected to the moving quark
might flip. Third, although at very low density quarks will
overwhelmingly belong to three-quark clusters, there is no guarantee
that this will remain true at higher densities; color-singlet clusters
may also be formed from 6-, 9-, $\ldots$ 3A-quark configurations (see
Fig.~\ref{Figure0}). Finally, the dynamically-induced residual
interaction between color-singlet clusters will be generated
exclusively through quark exchange, which induces a weak
intermediate-range attraction between clusters, and the Pauli
exclusion principle between quarks, which generates a strong
short-range repulsion.

As discussed early in this section, the strict demands imposed by QCD
on phenomenological models justifies the introduction of such a
complex many-body potential. While its exact functional form remains
uncertain, the requirements of quark confinement and cluster
separability are likely to depend on solving some type of quark
assignment problem. For simulations involving a large number of quarks
an efficient clustering algorithm is of utmost importance.  Indeed,
finding the optimal clustering of $N=3A$ quarks into $A$ color-singlet
objects requires searching among $(A!)^{2}$ configurations. Even for a
modest system containing only $A\!=\!10$ hadrons the number of
possible configurations already exceeds ten trillion! Clearly, a
``brute-force'' algorithm is impractical.  Moreover,
``three-dimensional stable matching problems'', such as the
three-quark assignment problem, have been shown to be
NP-complete~\cite{Ng91}. Thus, an efficient algorithm is unlikely to
exist. But for the version of the string-flip model adopted in this
work---where the clustering of quarks within color-singlet objects is
done pairwise---an efficient algorithm exists in the Hungarian method
for the weighted bipartite matching problem which finds the optimal
pairing in a time proportional to $A^{3}$~\cite{Mu80}.  Note that
while in this case the number of possible configurations grows``only''
as $A!$, a brute-force algorithm remains impractical.  Thus, without
such an efficient algorithm our simulations would be limited to a very
small number of quarks.

\subsection{The variational wave function}

We are interested in describing the evolution of the system with
baryon density. For that purpose we use a variational Monte-Carlo
approach based on a one-parameter wave function of the form:
\begin{equation}
 \Psi_{\lambda}(x)=e^{-\lambda V(x)}\Phi_{FG}(x) \;,
 \label{Psivar}
\end{equation}
where $\lambda$ is the variational parameter, $V(x)$ is the many-body
potential defined in Eq.~(\ref{MQpotential}), and $\Phi_{FG}(x)$ is
the Fermi-gas wave function.  This choice is motivated by QCD which
dictates that at low density, when the average inter-quark separation
is much larger than the confining scale, quarks should cluster into
three-quark color-singlet hadrons.  Thus, in the low-density regime
the potential between quarks in the same hadron is strong but, as the
interaction saturates within each cluster, the residual interaction
between hadrons is very weak. Hence, the system resembles a Fermi gas
of weakly interacting hadrons. It is the exponential term in the
variational wave function that becomes responsible for inducing the
clustering correlations. In contrast, in the high-density limit,
characterized now by an average inter-quark separation much smaller
than the confining scale, asymptotic freedom should take over. In this
regime the interactions between quarks are weak and the system
``dissolves'' into a free Fermi gas of quarks. As will be shown below,
the variational parameter evolves from a large (isolated-cluster)
value at low density all the way to zero at high density, as the only
remaining correlations between quarks are those induced by the Pauli
exclusion principle.

\subsubsection{Fermi-gas wave function}

To describe a non-interacting system of quarks a Fermi-gas wave
function, given in the form of a Slater determinant, is used for
each color-flavor combination of quarks. Each of these Slater
determinants is given by:
\begin{equation}
 \Phi_{FG}(x) = 
 \left| \matrix{
  \phi_{{\bf n}_{\hbox{\lower.3ex
                 \hbox{${\scriptscriptstyle 1}$}}}}({\bf x}_{1}) &
  \phi_{{\bf n}_{\hbox{\lower.3ex
                 \hbox{${\scriptscriptstyle 1}$}}}}({\bf x}_{2})  &
  \ldots &
  \phi_{{\bf n}_{\hbox{\lower.3ex
                 \hbox{${\scriptscriptstyle 1}$}}}}({\bf x}_{N}) \cr
  \phi_{{\bf n}_{\hbox{\lower.3ex
                 \hbox{${\scriptscriptstyle 2}$}}}}({\bf x}_{1}) &
  \phi_{{\bf n}_{\hbox{\lower.3ex
                 \hbox{${\scriptscriptstyle 2}$}}}}({\bf x}_{2})  &
  \ldots &
  \phi_{{\bf n}_{\hbox{\lower.3ex
                 \hbox{${\scriptscriptstyle 2}$}}}}({\bf x}_{N}) \cr
  \vdots & \vdots & \ddots & \vdots \cr
  \phi_{{\bf n}_{\hbox{\lower.3ex
                 \hbox{${\scriptscriptstyle N}$}}}}({\bf x}_{1}) &
  \phi_{{\bf n}_{\hbox{\lower.3ex
                 \hbox{${\scriptscriptstyle N}$}}}}({\bf x}_{2})  &
  \ldots &
  \phi_{{\bf n}_{\hbox{\lower.3ex
                 \hbox{${\scriptscriptstyle N}$}}}}({\bf x}_{N}) 
  } \right| \;.
 \label{PhiFG}
\end{equation}
where $\phi_{\bf n}({\bf x})$ represents a single-particle eigenstate
for a free particle in a box with quantum numbers ${\bf n}$ (see
below). This construction ensures that the wave function is totally 
antisymmetric under the exchange of identical quarks. To determine 
the single-particle wave functions we consider a single quark of mass 
$a$ confined to a three-dimensional box of side $a$ with antiperiodic 
boundary conditions. The energy of each single particle state is 
characterized by three integer quantum numbers ${\bf n}\!\equiv\!
(n_x, n_y, n_z)$:
\begin{equation} 
  E_{\bf n} = \frac{\pi^2}{2ma^2}(n_x^2+n_y^2+n_z^2) \;,
  \quad (n_{i}=1, 3, 5, \ldots) \;.
  \label{SPenergy} 
\end{equation}
Note that throughout this work we employ units in which 
$\hbar\!=\!c\!=\!1$. Each energy value, however, is at least 
eight-fold degenerate because there are even and odd solutions 
of the Schr\"odinger equation in each of the three spatial 
dimensions. Thus, a typical basis state is of the form:
\begin{equation}
  \phi_{n_x,n_y,n_z}^{+,+,-}({\bf x}) = 
  \cos\left(\frac{n_x\pi}{a}x\right)
  \cos\left(\frac{n_y\pi}{a}y\right)
  \sin\left(\frac{n_z\pi }{a}z\right) \;.
 \label{typical}
\end{equation}
In this way the system develops a ``shell structure'' with each 
shell holding, at least, eight quarks of each color-flavor 
combination. Let us illustrate how the shells are filled as the 
single-particle energy increases. Expressing the single-particle 
energies in units of ${\pi^2}/{2ma^2}$, the lowest shell has an 
energy of $3$ $(n_x\!=\!n_y\!=\!n_z\!=\!1)$ and is exactly 
eight-fold degenerate. The next shell, with an energy of $11$
($n_x\!=\!n_y\!=\!1,n_z\!=\!3$ and permutations) 
holds 24 quarks, and so on (for more details see 
Table~\ref{Table1}).

\subsubsection{Low-density limit}
\label{LDL}

In the low-density regime quarks are confined within weakly-interacting
color-singlet ({\em red+blue+green=white}) clusters. Further, although 
the model allows the existence of multi-quark configurations, the 
probability that color-singlet clusters with more than three quarks
are formed is vanishingly small~\cite{Ho92}. Finally, at these low 
densities the onset of strangeness is hindered by the large
strange-quark mass. Thus, in this limit the variational wave function
is exact, as we now show.

Consider a nucleon as a nonrelativistic system of three quarks of mass 
$m$ interacting via a confining potential assumed harmonic of spring
constant $k$
\begin{equation}
 H = \sum_{i=1}^{3}\frac{{\bf p}_{i}^{2}}{2m}  + 
     \sum_{i<j=1}^{3}
     \frac{1}{2}k({\bf r}_{i}-{\bf r}_{j})^{2} \;.
 \label{Hamilton}
\end{equation}
This is perhaps the simplest version of the surprisingly successful 
nonrelativistic quark model~\cite{Is79}. Introducing center-of-mass
and two relative coordinates
\begin{mathletters}
 \begin{eqnarray}
  {\bf R}_{CM} &=& \frac{1}{3} 
  ({\bf r}_{1}+{\bf r}_{2}+{\bf r}_{3}) \;, \\
  {\mbox{\boldmath $\xi$}_1} &=& \frac{1}{\sqrt{2}}  
  ({\bf r}_{1}-{\bf r}_{2}) \;, \\
  {\mbox{\boldmath $\xi$}_2} &=& \frac{1}{\sqrt{6}}  
  ({\bf r}_{1}+{\bf r}_{2}-2{\bf r}_{3}) \;,
 \end{eqnarray}
 \label{Jacobi}
\end{mathletters}
enables one to rewrite the above Hamiltonian as a system of two
uncoupled harmonic oscillators:
\begin{equation}
  H = \frac{P_{CM}^2}{6m} +
    \left(\frac{P_1^2}{2m}+\frac{3}{2}k\xi_1^2\right)  +
    \left(\frac{P_2^2}{2m}+\frac{3}{2}k\xi_2^2\right) \;.
 \label{HJacobi}
\end{equation}
The ground-state properties for the system are now easily inferred.
For example, the energy per quark is given by
\begin{equation}
 E_{0}/N - m = \omega = \sqrt{\frac{3k}{m}}
               \rightarrow \sqrt{3} \approx 1.732 \;.
 \label{EGS}
\end{equation}
where the arrow in the above expression is meant to represent 
the value of the energy in units in which $k\!=\!m\!=\!1$; this 
system of units is adopted henceforth. Further, up to an overall 
normalization factor, the ground-state wave function is also
easily computed; it is given by
\begin{equation}
 \Psi_{0}({\mbox{\boldmath $\xi$}_1},{\mbox{\boldmath $\xi$}_2})=
 \exp\left(-\frac{\xi_1^{2}+\xi_2^{2}}{2b^{2}}\right) =
 \exp\left(-\frac{v({\mbox{\boldmath $\xi$}_1},
                    {\mbox{\boldmath $\xi$}_2})}{3kb^{2}}\right) \;,
 \label{PsiGS}
\end{equation}
where the oscillator-length parameter 
$b\!\equiv\!(3km)^{-1/4}\!\rightarrow\!3^{-1/4}$ has been introduced.
Note that we have used the fact that the above exponent is
proportional to the potential energy of the baryon [see
Eq.~(\ref{HJacobi})] to write the second expression for the
ground-state wave function.  This expression suggests that in the
limit of very low density the variational wave function defined in
Eq.~(\ref{Psivar}) becomes exact provided:
\begin{equation}
 \lambda = \frac{1}{3kb^{2}} \longrightarrow 
 \frac{1}{\sqrt{3}} \approx 0.577 \;.
 \label{lambda0}
\end{equation}
Note that in the low-density limit the Fermi-gas component of
the wave function is not important, as the average separation 
between quarks of the same color-flavor combination is much 
larger than the ``Pauli hole''.

\subsubsection{High-density limit}
\label{HDL}
The variational wave function is also exact in the limit of very high
density. In this asymptotically-free regime the interaction between
quarks is negligible so the only remaining correlations among them are
those generated by the Pauli-exclusion principle.  Thus, the system is
described by a Fermi-gas wave function which represents the
$\lambda\rightarrow 0$ limit of the variational wave function given in
Eq.~(\ref{Psivar}).

A Fermi-gas description is useful in establishing a baseline 
against which more sophisticated models may be compared. In
addition, all observables that will be presented here can be 
computed analytically in the Fermi-gas limit. Hence let us 
start by computing the transition density to strange matter. 
This critical density is obtained by requiring that the chemical 
potential for the light ($u$ and $d$) quarks be identical to the 
strange-quark mass ($M$). That is,
\begin{equation}
  m + \frac{k_{Fc}^{2}}{2m} = M \;\; {\rm or} \;\;
  k_{Fc} = \sqrt{2m(M-m)} \rightarrow 1.095   \;.
 \label{kfcrit}
\end{equation}
Note that we have adopted a value of $M/m\!=\!1.6$ for the
heavy-to-light mass ratio. Adopting a constituent light-quark
mass of $m\!=\!300$~MeV, the transition density in physical 
units corresponds (with $N_{u}\!\equiv\!N_{d}$) to
\begin{equation}
  k_{Fc} = 328.634~{\rm MeV} \;\; {\rm or} \;\;
  \rho_{c}=\frac{k_{Fc}^{3}}{\pi^{2}} = 
  0.468~{\rm fm}^{-3} \;.
 \label{rhocrit}
\end{equation}
Note that spin degrees of freedom, or spin-dependent interactions, 
have yet to be included in the model. 

Let us proceed to evaluate the equation of state and the 
strangeness-to-baryon ratio as a function of density. First, 
however, we introduce the following definitions: 
\begin{equation}
  \rho=\frac{k_{F}^{3}}{\pi^{2}} \;\; {\rm and} \;\;
  \sigma=\frac{N_{s}}{N}=1-2\frac{N_{u}}{N} \;.
 \label{Defs}
\end{equation}
In terms of these relations the Fermi momenta become
\begin{equation}
 k_{F}^{u} = (1-\sigma)^{1/3} k_{F} \;\; {\rm and} \;\;
 k_{F}^{s} = (2\sigma)^{1/3} k_{F}  \;.
 \label{kFDefs}
\end{equation}
Moreover, the total energy-per-particle of the system --- at 
fixed density ($\rho$) and strangeness-per-quark ($\sigma$) 
--- may now be easily computed. We obtain
\begin{equation}
  \frac{T_{FG}(\rho,\sigma)}{N} 
          = (1-\sigma)m + \sigma M
          + \frac{3k_{F}^{2}}{10m}(1-\sigma)^{5/3}
          + \frac{3k_{F}^{2}}{20M}(2\sigma)^{5/3} \;.
 \label{TFG}
\end{equation}
The strangeness-per-quark is now determined by demanding that the 
total energy of the system be minimized at fixed density. That is,
\begin{equation}
  \left(\frac{\partial T_{FG} /N}{\partial \sigma}\right)  =
  \left[M+\frac{k_{F}^{2}}{2M}(2\sigma)^{2/3}\right]       -
  \left[m+\frac{k_{F}^{2}}{2m}(1-\sigma)^{2/3}\right]      =0 \;.
 \label{dTFG}
\end{equation}
This equation reflects the condition for chemical equilibrium:
the chemical potential for both species of quarks (light and heavy) 
must be equal. In particular, at $\sigma\!\equiv\!0$, which 
corresponds to the onset of the transition to strange-quark matter, 
one recovers the critical density computed in Eq.~(\ref{kfcrit}).

Another useful observable to characterize the transition to quark
matter is the two-body correlation function~\cite{Fe71}
\begin{equation}
 \rho{\hbox{\lower 3pt \hbox{${_2}$}}}({\bf r}) =
  \sum_{\alpha\beta} 
  \langle\Psi_{0}|
   \hat{\psi}_{\alpha}^{\dagger}({\bf r})
   \hat{\psi}_{\beta}^{\dagger}({\bf 0})
   \hat{\psi}_{\beta}({\bf 0})
   \hat{\psi}_{\alpha}({\bf r})|\Psi_{0}\rangle \;,
 \label{rho2}
\end{equation}
where $\alpha$ (and $\beta$) denotes the collection of all internal 
quantum numbers, such as color and flavor. The two-body correlation 
function measures the probability that two given quarks be separated 
by a distance ${\bf r}$. In the Fermi-gas limit, where the ground-state
wave function may be written as a Slater determinant, the two-body 
correlation function may be readily evaluated. We obtain,
\begin{equation}
  g{\hbox{\lower 3pt \hbox{${_2}$}}}({\bf r}) \equiv
  \frac{\rho{\hbox{\lower 3pt \hbox{${_2}$}}}({\bf r})}
       {\rho^{2}} = 1-\frac{1}{3}
  \left(\frac{3j_{1}(k_{F}r)}{k_{F}r}\right)^{2} \;,
 \label{gtwo}
\end{equation}
where the spherical Bessel function is given by
\begin{equation}
  j_{1}(x)=\left(\frac{\sin x}{x^2} - \frac{\cos x}{x}\right)
           \mathop{\longrightarrow}_{x\rightarrow 0}
           \; \frac{x}{3} \;.
 \label{jone}
\end{equation}
Note that $g_{2}$ has been normalized to one at large distances.
Moreover, the correlation function between two quarks of the same 
flavor develops a ``hole'' at the origin as a consequence of the 
Pauli exclusion principle. Yet the Pauli suppression is not complete 
[$g_{2}(0) \ne 0$] because in our model quarks of the same flavor 
carry an additional color quantum number.

\subsubsection{Variational Monte-Carlo simulations}

The variational nature of the simulation suggests that the expectation
value of the Hamiltonian, Eq.~(\ref{HamiltonV}), will have to be minimized with respect to the
variational parameter $\lambda$ introduced in Eq.~(\ref{Psivar}). That
is,
\begin{equation}
  \frac{\partial E(\lambda)}{\partial\lambda}=0
  \;\; {\rm where} \;\;
  E(\lambda) = \langle\Psi_{\lambda}|H|\Psi_{\lambda}\rangle \;.
 \label{ELambda1}
\end{equation}
Note that the expectation value of the energy as a function of 
$\lambda$ will have to be computed for all densities and for a 
variety of strangeness-to-baryon ratios. The computational 
demands imposed on such a calculation are formidable, indeed. 
Yet, the structure of the variational wave function entails
some simplifications. For example, the expectation value of 
the kinetic-energy operator may be simplified through an 
integration by parts. That is,
\begin{equation}
  \langle\Psi_{\lambda}|T|\Psi_{\lambda}\rangle =
  T_{FG} + 2\lambda^2 \langle W \rangle_{\lambda} \;.
 \label{Tlambda}
\end{equation}
where $T_{FG}$ is the kinetic energy of a ($\lambda\!=\!0$) free 
Fermi gas and $\langle W \rangle_{\lambda}$ reflects the increase 
in the kinetic energy of the system relative to the Fermi-gas 
estimate due to clustering correlations. It is given by
\begin{equation}
    W  = \sum_{n=1}^{N}\frac{1}{m_{n}}
    \Big({\bf x}_{n}-{\bf y}_{n}\Big)^{2} \;,
 \label{Wpotent}
\end{equation}
where the sum is over all quarks in the system and ${\bf y}_{n}$ 
represents the average position of the two quarks connected to
the $n_{\rm th}$ quark. Note that in the limit that only 
three-quark clusters are formed and $m_{n}\!=\!1$, such as in the 
low-density limit, then $W \!=\! 3V/2$. Now using Eq.~(\ref{Tlambda}) 
we obtain the following simple form for the expectation value of 
the total energy of the system:
\begin{equation}
  E(\lambda) = T_{FG} + 2\lambda^2 
               \langle W \rangle_{\lambda}
             + \langle V \rangle_{\lambda} \;.
 \label{ELambda2}
\end{equation}
This form is particularly simple because the two functions that
remain to be evaluated ($V$ and $W$) are local; thus, their 
expectation values may be easily computed via Monte-Carlo methods. 
To do so we use the well-know Metropolis method~\cite{Me53}. 

The Metropolis algorithm is based on a Markov process that 
generates events (or configurations) stochastically. The
Markov chain is created sequentially from knowledge of
only the current configuration. That is, the $(m+1)_{\rm th}$ 
configuration in the chain is generated stochastically using 
only the $m_{\rm th}$ configuration; no information about the 
$(m-1)_{\rm th}$ event is required at all. We illustrate briefly
the method with the evaluation of the expectation value of the 
potential energy
\begin{equation}
  \langle V \rangle_{\lambda} = 
  \langle\Psi_{\lambda}|V|\Psi_{\lambda}\rangle =
  \int V({\bf x}_{1},\ldots,{\bf x}_{N})
       \Psi_{\lambda}^{2}({\bf x}_{1},\ldots,{\bf x}_{N})\,
        d{\bf x}_1 \ldots d{\bf x}_N \;.
 \label{ExpV}
\end{equation}
Although in writing this expression we have assumed a normalized
wave function, it is not necessary for the wave function to be
normalized when implementing the Metropolis algorithm. Note that
for a system of $N\!=\!120$~quarks, as we have used in some of
our simulation, computing the above expectation value requires 
the evaluation of a 360-dimensional integral! The Metropolis
algorithm ensures that the desired probability distribution,
$\Psi_{\lambda}^{2}$ in our case, is approached asymptotically.
The main idea of the method is not to evaluate the integrand at 
every one of the quadrature points, an impossible task indeed, 
but rather at only a relatively small representative 
sampling~\cite{Ko86}. That is, the expectation value of the
potential energy becomes
\begin{equation}
  \langle V \rangle_{\lambda} =
  \lim_{M\rightarrow\infty} \frac{1}{M}
  \sum_{m=1}^{M} 
  V\left({\bf x}_{1}^{(m)},\ldots,{\bf x}_{N}^{(m)}\right) \;,
 \label{VMC}
\end{equation}
where the sequence of $M$ configurations are distributed according
to $\Psi_{\lambda}^{2}$.

\section{Results}
\label{results}

\begin{figure}
\centerline{\epsfig{file=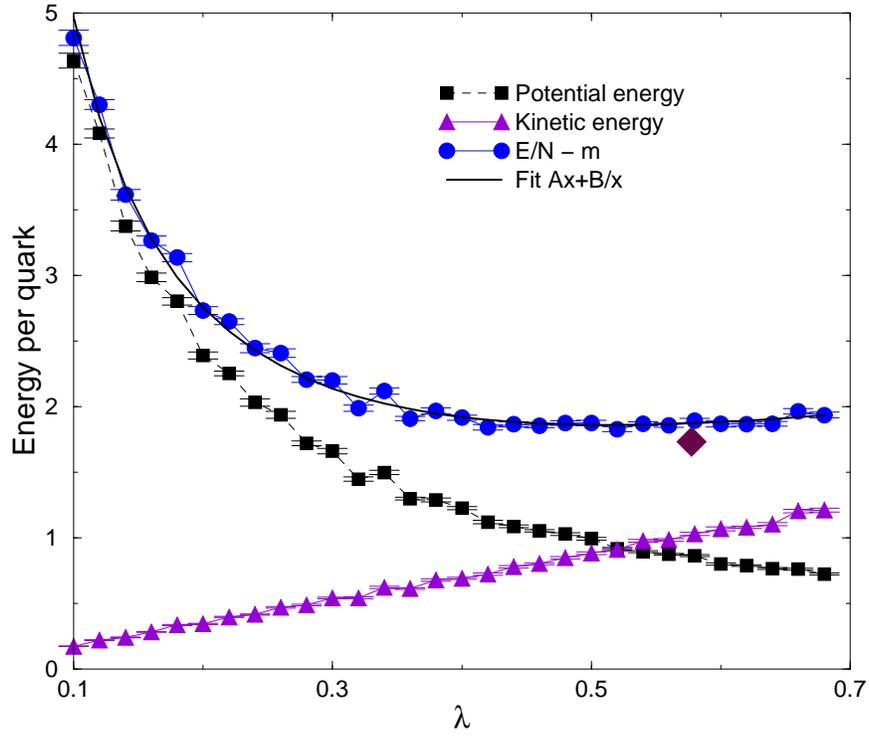,angle=-90,width=4.5in}}
\vspace{0.0in}
\caption{Low-density behavior of the kinetic, potential, and total 
         energy per quark as a function of the variational parameter. 
	 The diamond indicates the energy/quark and variational 
	 parameter for an isolated nucleon. The density of the
	 system is $\rho/\rho_{c}\!\simeq\!4\!\times\!10^{-3}$.} 
\label{ela0}
\end{figure}

\begin{figure}
\centerline{\epsfig{file=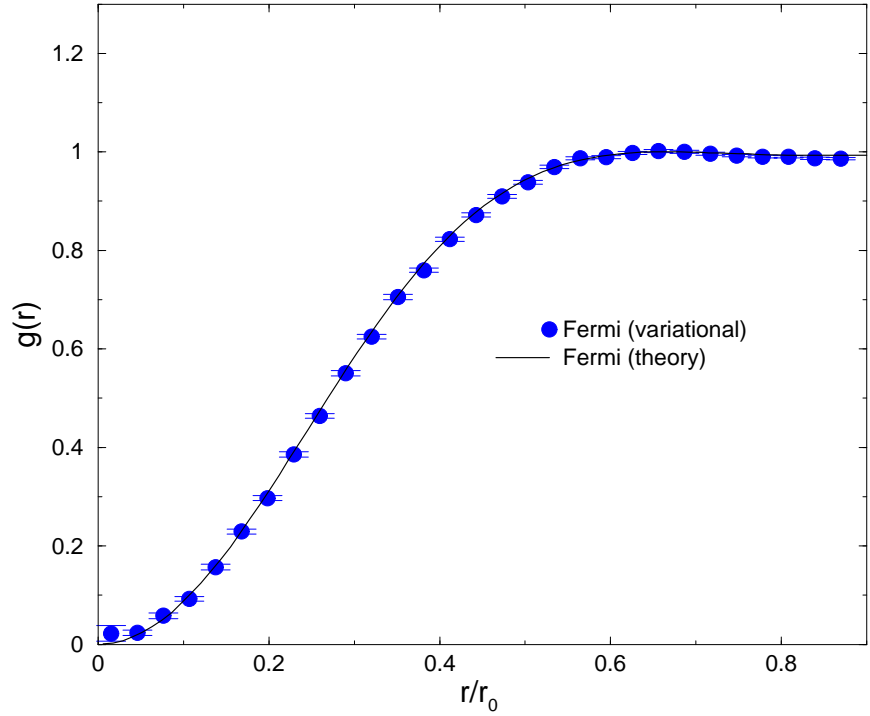,angle=-90,width=4.5in}}
\vspace{0.0in}
\caption{Two-body correlation function between identical quarks.
         The circles represent the results from the Monte-Carlo
	 simulation while the solid line is the theoretical 
	 prediction. The density of the system is 
	 $\rho/\rho_{c}\!\simeq\!45$.} 
\label{cluster}
\end{figure}

As a test of the formalism and to illustrate how the variational
approach becomes exact in the low- and high-density limits, we display
in Figs.~\ref{ela0} and~\ref{cluster} the ground-state energy of the
system and its two-body correlation function, respectively. All simulations performed in this work were done using 120 quarks. At
very-low density the system resembles a non-interacting gas of
nucleons with an energy-per-nucleon and variational parameter
identical to the single-nucleon values given in Sec.~\ref{LDL}.  That
this is indeed the case is shown In Fig.~\ref{ela0}. Here the kinetic,
potential, and total energy of the system are plotted as a function of
the variational parameter for a density of
$\rho/\rho_{c}\!\simeq\!4\!\times\!10^{-3}$.  (Note that the
light-quark rest mass has been subtracted out). As in the case of an
isolated nucleon, the plot reflects the competition between the
kinetic energy, which tends to diffuse the wave function away from the
origin (favors a small value of $\lambda$) and the potential energy,
which attempts to concentrate the wave function at the origin (favors
a large value of $\lambda$).  A compromise is reached, in accordance
to the virial theorem, at the point at which the kinetic and potential
energies are equal; that is, at
$\lambda\!\simeq\!\lambda_{0}\!=\!1/\sqrt3$.

In the opposite high-density limit the system is expected to 
evolve into a collection of non-interacting quarks. Thus, it
should display no correlations other than those generated by 
the Pauli exclusion principle. A simple way to test this 
assertion is by computing the two-body correlation function 
between identical quarks. If the system has indeed evolved 
into a Fermi gas of quarks, the two-body correlation function 
should become identical to the one given in Eq.~(\ref{gtwo}), 
with the ``color prefactor'' of $1/3$ set up to one. This 
analytic expression is plotted in Fig.~\ref{cluster} (solid 
line) for a the very large density of 
$\rho/\rho_{c}\!\simeq\!45$; the agreement with the numerical 
simulations (filled circles) is excellent indeed.

\begin{figure}
\centerline{\epsfig{file=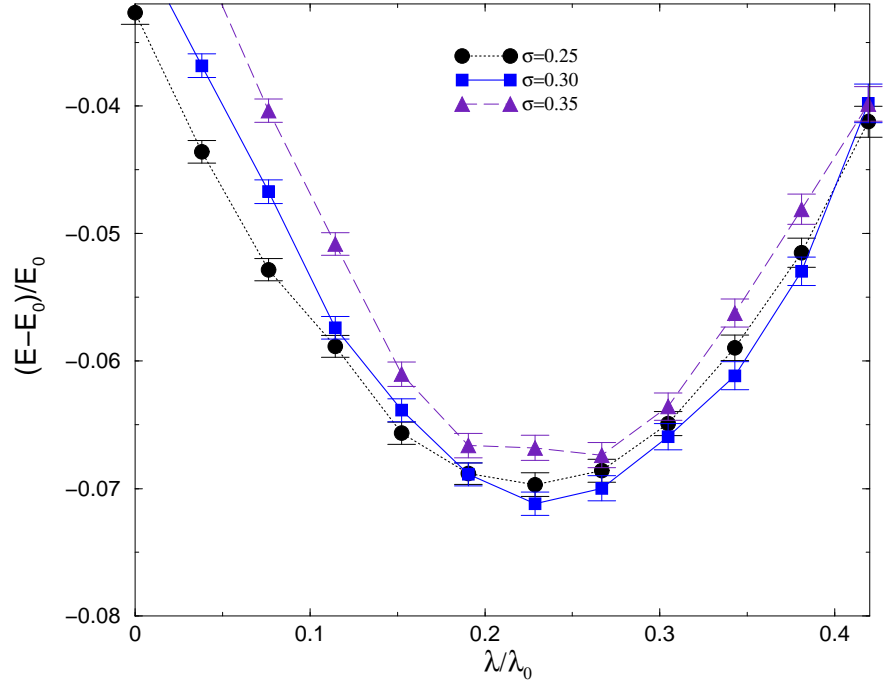,angle=-90,width=4.5in}}
\vspace{+0.0in}
\caption{Energy-per-quark as a function of the variational
	 parameter for three values of the strangeness-to-quark
	 ratio. The number of quarks in the simulation
	 is $N\!=\!120$ and the density of the system is fixed
	 at $\rho/\rho_{c}\!=\!2.18$.}
\label{min}
\end{figure}

The determination of the variational parameter is a non-trivial
computational task. First, one must select the quark density of the
system $\rho$. For that particular density one then fixes the
strangeness-to-quark ratio $\sigma$. Having fixed these two quantities
one then proceeds to compute the energy of the system as a function of
the variational parameter using the Monte-Carlo methods described
earlier in the text. The outcome of such a calculation is one of the
three curves shown in Fig.~\ref{min}. Note that in this plot, and
throughout the remainder of the paper, $E_{0}$ and $\lambda_{0}$
represent the energy and variational parameter of an isolated nucleon
(see Sec.~\ref{LDL}). In order to generate the remaining curves in the
figure one must change the strangeness-to-quark ratio while
maintaining the baryon density fixed. Once a minimum in the energy is
identified, the energy-per-quark, strangeness-to-quark ratio, and
variational parameter for that particular value of the baryon
density are determined.  Figure~\ref{min} shows the outcome of this
lengthy procedure for the particular case of
$\rho/\rho_{c}\!=\!2.18$. This procedure must then be repeated for all
quark densities.

\begin{figure}
\centerline{\epsfig{file=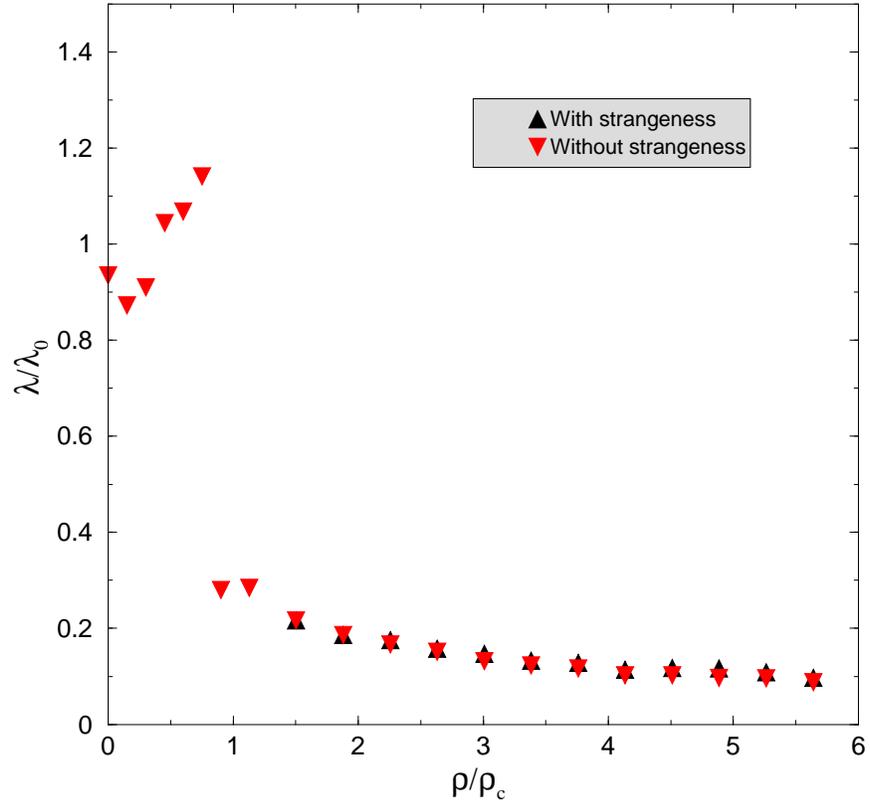,angle=-90,width=4.5in}}
\vspace{0.0in}
\caption{Variational parameter $\lambda$ as a function of 
         density for systems with (up-triangles) and without 
         (down-triangles) strangeness. The sharp drop signals 
	 the transition from three- to multi-quark configurations.}
\label{dl}
\end{figure}

The density dependence of the variational parameter $\lambda$ is
displayed in Fig.~\ref{dl}. The behavior of this quantity with density
is interesting as $\lambda^{-1/2}$ may be regarded as the length-scale
for quark confinement. At very low densities there is a drop in the
value of $\lambda$ indicating that the length-scale for quark
confinement has increased in the medium; this is reminiscent of the
``nucleon swelling'' first observed in the deep-inelastic scattering
by the EMC collaboration. Yet, this is followed by a relative steep
increase in the value of $\lambda$ suggesting that the system is
favoring the formation of highly tight nucleons. This unexpected
behavior was reported in Ref.~\cite{Ho92} for the one-flavor
model. Eventually the system must make the transition into the
Fermi-gas domain ($\lambda\rightarrow 0$); this is accomplished
through an abrupt drop at a quark density of about
$\rho/\rho_{c}\!=\!0.82$.  This density represents the transition 
from three-quark clusters into multi-quark
configurations~\cite{Ho92}. Fig.~\ref{dl} also indicates that while
the presence of strange quarks has a minimal effect in the
density-dependence of $\lambda$, clustering correlation remain
non-negligible well into the strange-matter domain.

\begin{figure}
\centerline{\epsfig{file=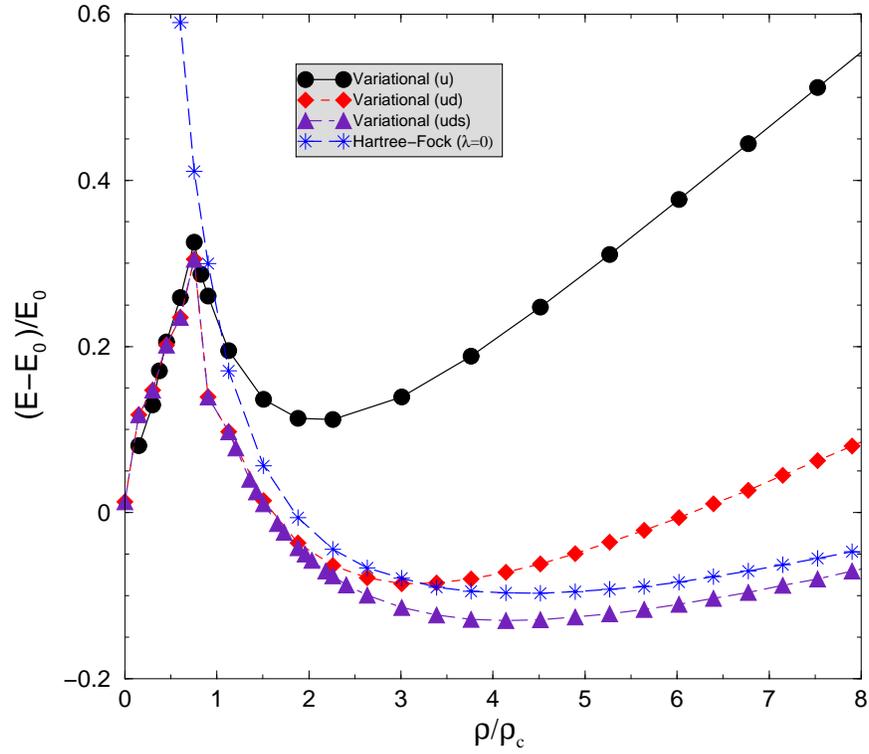,angle=-90,width=4.5in}}
\vspace{0.0in}
\caption{Energy per quark as a function of density for one-, two- 
	 and three-flavor systems as a function of density. The
	 Hartree-Fock ($\lambda\!=\!0$) energy is also shown 
	 to illustrate the effect of clustering correlations.
	 In this model three-flavor strange-quark matter is 
	 absolutely bound.}
\label{de123}
\end{figure}

In Fig.~\ref{de123} we show the energy-per-quark as a function of the
density obtained with the variational Monte-Carlo approach.  Various
calculations are displayed in the figure. The one-flavor calculation
of Ref.~\cite{Ho92} is reproduced with the filled circles. While the
equation of state displays a minimum in the strange-quark matter
domain, the minimum is local so the system becomes unstable against
the break-up into isolated three-quark nucleons. However, flavor
degeneracy stabilizes the strange-matter phase in these
models. Indeed, in both the two- (diamonds) and three-flavor
(triangles) cases quark-matter is absolutely stable.  Note that at the
point at which the variational parameter drops discontinuously (see
Fig.~\ref{dl}) the energy-per-quark reaches its maximum value and then
drops rapidly with density, although in a continuous manner. This
density ($\rho/\rho_{c}\!=\!0.82$) represents the transition from
three- to multi-quark configurations.  At the higher density of
$\rho/\rho_{c}\approx 1.2$ two- and three-flavor matter separate in
the plot indicating the transition into strange-quark matter. Finally,
we have included a Hartree-Fock ($\lambda\!=\!0$) calculation to
illustrate the importance of clustering correlations, even in the
strange-quark domain. It is worth mentioning that non-appreciable
finite size effects were observed for simulations using 90 quarks.

\begin{figure}
\centerline{\epsfig{file=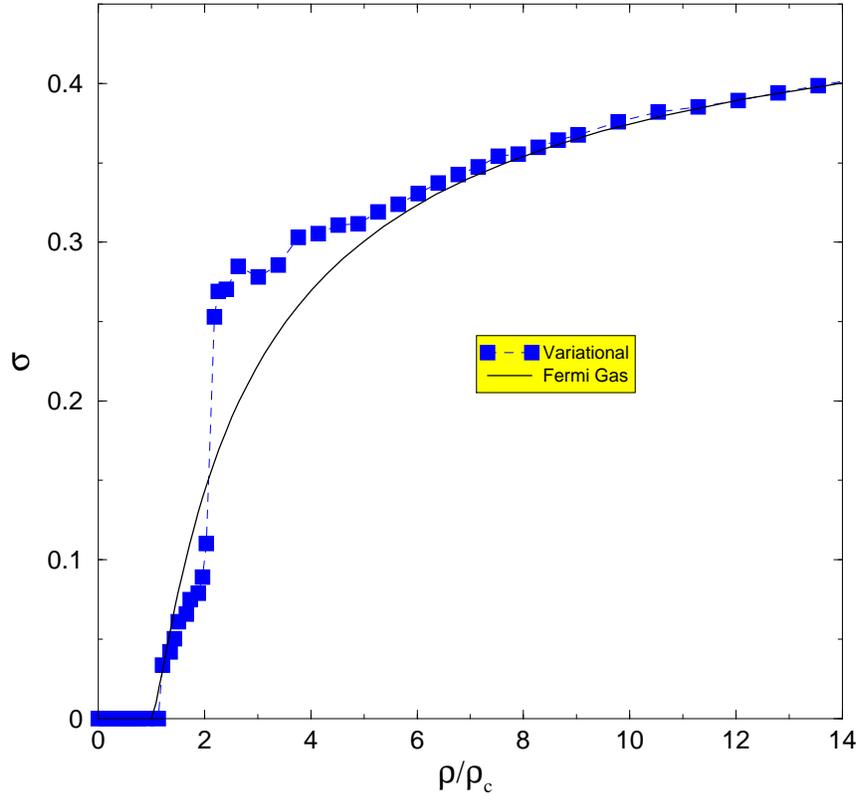,angle=-90,width=4.5in}}
\vspace{0.0in}
\caption{Strangeness-to-quark ratio as a function of density for the
	 variational calculation (filled squares) and a simple 
	 Fermi-gas estimate (solid line). Note the discontinuity of 
	 the transition at $\rho/\rho_{c}\!\approx\!2.2.$}
\label{sgm}
\end{figure}

\begin{figure}
\centerline{\epsfig{file=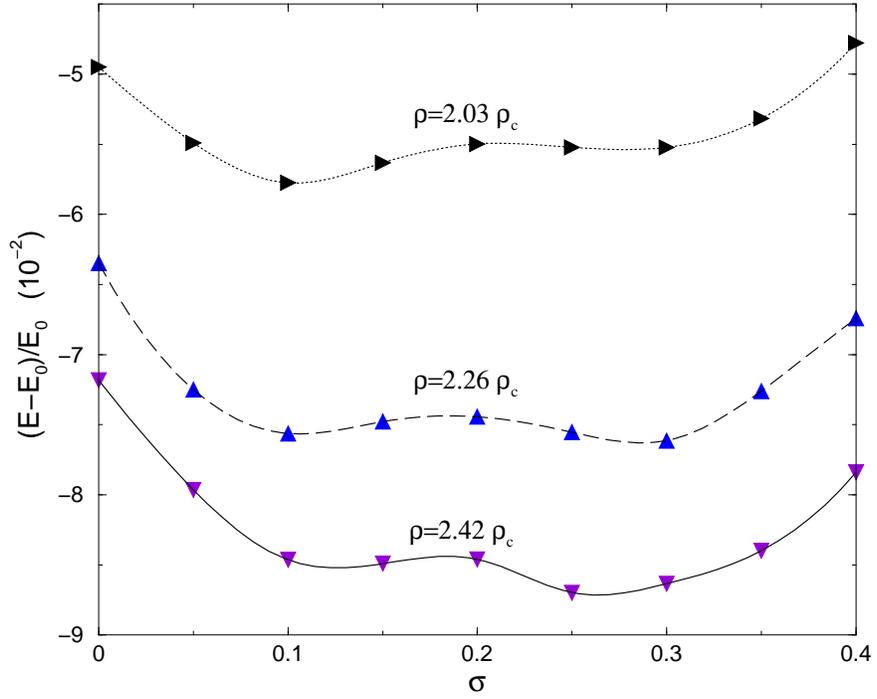,angle=-90,width=4.5in}}
\vspace{0.0in}
\caption{Energy-per-quark as a function of strangeness-to-quark ratio
         for three values of the quark density. In all three cases the 
	 variational parameter has been fixed at its optimal value.
         Note the competition between the two minima, particularly in
         the $\rho/\rho_{c}\!\approx\!2.26$ case.}
 \label{minco}
\end{figure}

In Fig.~\ref{sgm} the strange-quark content of hadronic matter is
plotted as a function of the quark density. The solid line represents
the analytic result obtained for a free Fermi gas of quarks with a
ratio of strange-to-light quark masses of $M/m\!=\!1.6$. The results
of the Monte-Carlo simulations are displayed with the filled squares.
Note that the transition to strange matter is slightly delayed
relative to the Fermi-gas predictions; this represents a small
quantitative change that is in agreement with the one-dimensional
predictions of Ref.~\cite{Mo99}.  A much more interesting change
happens at a density of $\rho/\rho_{c}\!\approx\!2.2$: a second
minimum develops well inside the strange-matter phase. The evolution
with density of this second minima is illustrated in Fig.~\ref{minco}.
While the competing (large-$\sigma$) minimum is shallow at the the
lowest density shown in the figure, it becomes the absolute ground
state of the system at the highest density.  The change from shallow
to deep is accompanied by a discontinuous phase transition. At higher
densities the variational and theoretical results match as expected.

\section{Conclusions}
\label{conclusions}
Because of its intrinsic interest as the possible ground state of
hadronic matter as well as its relevance to the dynamics of neutron
stars, strange-quark matter has been modeled directly in terms of its
quark constituents. A three-flavor, three-color string-flip model that
confines quarks within individual color-neutral clusters, yet allows
the clusters to separate without generating unphysical long-range
forces, was used to compute strange-matter observables.  Perhaps the
greatest virtue of the model is that the transition from nuclear
matter---where quarks are confined within color-singlet hadrons---to
quark matter---where quarks are free to roam through the simulation
volume---is dynamical; there is no need to introduce ad-hoc parameters
to characterize the transition. Indeed, at very low densities the
quarks cluster dynamically into (three-quark) nucleons having
properties identical to those in free space. As the density increases
the clusters dissolve, again dynamically, into a uniform Fermi gas of
quarks. This is in contrast to most of the models available today that
use either nucleon and hyperons, even at very high densities, or
quarks in a bag as the fundamental degrees of freedom.

After testing our variational Monte-Carlo approach in the low- and
high-density limits, where the approach becomes exact, we proceeded to
compute the variational parameter, the equation of state, and the
strangeness content of strange matter. We were able to identify two
phase transitions in the model. The first one, previously reported in
the one-flavor simulations of Ref.~\cite{Ho92}, is characterized by a
transition from three- to multi-quark cluster configurations.  The
variational parameter, which is directly related to the length scale
for quark confinement, jumps discontinuously at the transition.  A
transition to strange-matter was also identified at a higher density.
The onset for the transition was delayed slightly relative to the
predictions of a free Fermi-gas estimate.  Much more interesting,
however, was the development of an additional discontinuous phase
transition well into the strange-matter domain. This phase transition
was characterized by the competition between two ground states; one
with a low and one with a high strange-quark content. We should point
out that the effect of clustering correlations remained important even
at the highest transition density. Also interesting was the behavior
of the equation of state as a function of the flavor-content of
hadronic matter. For a simplified one-flavor model strange matter
saturates but is unstable against the break-up into isolated
nucleons. Yet as the flavor-content of the model was increased,
strange matter became absolutely stable.

It is too early to predict how far we will be able to push this
model. Ultimately, one would hope to compute a realistic equation of
state that could be used as input in the computation of masses and
radii of neutron stars.  Unfortunately, and in spite of the
considerable effort devoted to this endeavor, several crucial
refinements remain to be added. Perhaps the most important one is
related to the lack of binding in the low-density nuclear phase. In
the most pristine form of the model quark exchange is the unique
source of attraction. Clearly, this is insufficient to bind nuclear
matter.  Perhaps adding the missing physics associated with the
long-range pionic tail might help solve this problem. Further, while
the present version of the model incorporated flavor and color degrees
of freedom for the first time, spin and spin-dependent interactions
remain to be included.  Incorporating this extra degree of freedom in
the calculations, while avoiding an over-proliferation of baryonic
states, remains a difficult challenge.  Finally, it is unrealistic to
expect that a non-relativistic description will remain valid in the
high-density domain. This problem can be solved, at least in part, by
using a relativistic form for the kinetic energy of a free Fermi gas.
In spite of all these challenges we believe that the string-flip model
used here represents a sound starting point for the description of
this complicated many-body system. In particular, we are convinced
that ultimately some form of quark-assignment problem will have to be
solved in order to account simultaneously for quark confinement and
cluster separability.

\acknowledgements
\medskip
This work was supported in part by the United State Department of
Energy under Contract No.DE-FG05-92ER40750. GTS thanks CONACYT, 
M\'exico for a postdoctoral fellowship.

\begin{table}[h]
\caption{Shell structure of a Fermi-gas (Slater) 
	 determinant for fermions of mass $m$ 
	 occupying a three-dimensional box of size 
	 $a$ with antiperiodic boundary conditions. 
	 All energies are measured in units of
         ${\pi^2}/{2ma^2}$.}
 \begin{tabular}{cccccc}
 $n_{x}$ &  $n_{y}$ &  $n_{z}$ & Energy & 
 Total Occupancy \\
 \hline \hline
  1  &  1  &  1  &   3  &     [8]  \\
 \hline
  1  &  1  &  3  &  11  &     [16]  \\
  1  &  3  &  1  &  11  &     [24]  \\
  3  &  1  &  1  &  11  &     [32]  \\
 \hline
  3  &  3  &  1  &  19  &     [40]  \\
  3  &  1  &  3  &  19  &     [48]  \\
  1  &  3  &  3  &  19  &     [56]  \\
 \hline
  3  &  3  &  3  &  27  &     [64]  \\
  1  &  1  &  5  &  27  &     [72]  \\
  1  &  5  &  1  &  27  &     [80]  \\
  5  &  1  &  1  &  27  &   \ [88]  
  \label{Table1}
 \end{tabular}
\end{table}

\vfill\eject
\vfill\eject
\vfill\eject
\vfill\eject
\vfill\eject
\vfill\eject
\vfill\eject

\vfill\eject

\end{document}